\documentclass[twocolumn,preprintnumbers,amsmath,amssymb]{revtex4}
\usepackage{graphicx}
\usepackage{dcolumn}
\usepackage{xcolor}

\usepackage{bm}

\begin{document}

\title{Microscopic theory of photonic band gaps in optical lattices}

\author{M. Samoylova$^{a}$}
\author{N. Piovella$^{a}\footnote{Corresponding author. Tel:+39 02 50317266. E-mail address: nicola.piovella@unimi.it}$}
\author{R. Bachelard$^{b}$}
\author{Ph.W. Courteille$^{b}$}
\affiliation{$^{a}$Dipartimento di Fisica, Universit\`a degli Studi di Milano,Via Celoria 16, I-20133 Milano, Italy\\
$^{b}$Instituto de F\'isica de S\~ao Carlos, Universidade de S\~ao Paulo, 13560-970 S\~ao Carlos, SP, Brazil}

\date{\today}

\begin{abstract}
\textbf{Abstract}. We propose a microscopic model to describe the scattering of light by atoms in optical lattices.
The model is shown to efficiently capture Bragg scattering, spontaneous emission and photonic band gaps.
A connection to the transfer matrix formalism is established in the limit of a one-dimensional optical lattice,
and we find the two theories to yield results in good agreement. The advantage of the microscopic model is, however,
that it suits better for studies of finite-size and disorder effects.


\end{abstract}

\maketitle

\section{Introduction.}

When an electromagnetic wave is sent into an atomic cloud, the interference of the radiation fields emitted by every atom gives
rise to cooperative scattering. In disordered systems this interference phenomenon was first described by Dicke who evidenced
the superradiant emission of the cloud~\cite{Dicke1954}. Later, other striking features linked to cooperative scattering have
been observed, such as the collective Lamb shift~\cite{Friedberg1973,Scully2010,Keaveney2012} and a reduction of the radiation
pressure force~\cite{Bienaime2010,Courteille2010}. The search for the localization of light by the disorder itself is still
ongoing~\cite{Wiersma1997,Conti2008}.

The hallmark of the photonic properties of \emph{ordered atomic ensembles}, such as optical lattices, is the formation of a
band structure similar to those encountered in photonic crystals. Photonic band gaps (PBG) have been predicted in one-dimensional
arrays of atomic clouds using the transfer matrix (TM) technique~\cite{Deutsch1995}. This approach, which relies on the description
of an atomic ensemble as a continuous dielectric with a very large transverse size, describes well the situation of recent
experiments~\cite{Slama2005,Slama2005a,Slama2006}, which culminated in the first observation of a PBG in a one-dimensional
optical lattice~\cite{Schilke2011,Schilke2012}.

In the case of three-dimensional optical lattices, the Bloch-Floquet model has been used to calculate the propagation of
electromagnetic modes in Fourier space and identify omnidirectional PBGs in certain geometries assuming infinite and perfectly
periodic lattices~\cite{Antezza2009,DeshuiYu2011}. Nevertheless, omnidirectional PBGs remain to be observed experimentally.

\section{Microscopic model.}

In this paper we propose a microscopic model of cooperative scattering from an ordered atomic gas, treating the atoms as point-like
scatterers interacting with light via an internal resonance. We show that this model is able to describe the opening of a forbidden
photonic band due to multiple reflection of light between adjacent lattice sites. We support our assertion in two ways.
Using numerical simulations of the microscopic model we find that Bragg scattering and PBGs arise in our system.
We also demonstrate that under a coarse-graining hypothesis and in the limit of a one-dimensional optical lattice, the microscopic model
boils down to the TM formalism used, e.g., in Refs.~\cite{Deutsch1995,Slama2006,Schilke2011}.

It must be highlighted that our microscopic model does not contain
the limitations of the above-mentioned other techniques. In
particular, it does not reduce the atomic layers to a smooth
dielectric, nor does it assume the atomic cloud to be perfectly
periodic or infinite in any direction. It is thus notably suited
to study the role of the disorder and finite-size effects on
photonic bands.

The collective light scattering by an atomic ensemble is described by the following coupled equations~\cite{Bachelard2011, Bienaime2011, Morice1995}:
\begin{equation}
\label{eq:stationary solution}
    \left(i\Delta_0-\frac{\Gamma}{2}\right)\beta_j=\frac{i\wp}{2\hbar}E_0(\textbf{r}_j)+\frac{\Gamma}{2}\sum_{k\ne
        j}\frac{\exp(ik_0|\textbf{r}_j-\textbf{r}_k|)}{ik_0|\textbf{r}_j-\textbf{r}_k|}\beta_k~
\end{equation}
where $\textbf{r}_j$ is the position of the $j$th atom and $\beta_j$ is the excitation amplitude of its dipole. The first term on the right-hand
side of Eq.~(\ref{eq:stationary solution}) corresponds to the field $E_0(\textbf{r})$ of the incident laser beam, whereas the last term
characterizes the radiation from all other atoms. $\Delta_{0}$ is the detuning of the incident laser with respect to the atomic transition,
$\Gamma$ is the single atom spontaneous decay rate and $\wp$ is the electric dipole matrix element.

\begin{figure}[t]
\centerline{{\includegraphics[width=7cm]{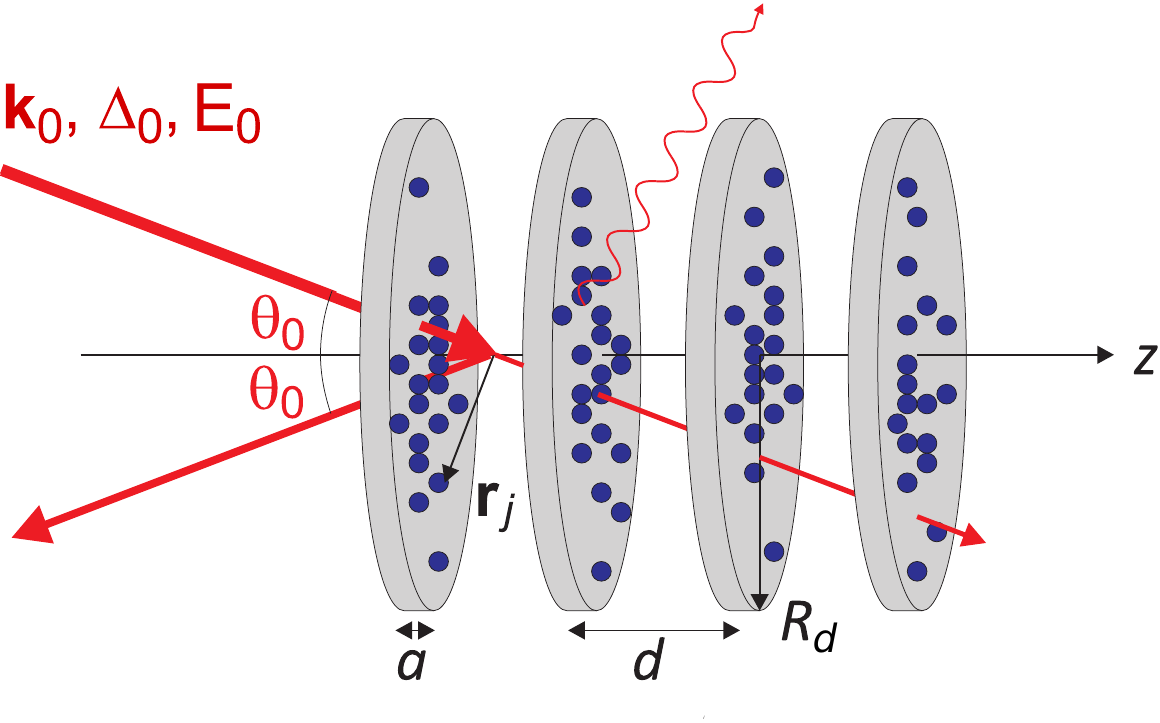}}}
\caption{(color online)  Experimental setup: An array of disks
randomly filled with atoms is irradiated from the side by a probe
beam incident under an angle $\theta_0$. The system can be
considered as one-dimensional assuming that $a,d\ll R_d$, where
$a$ and $R_d$ are the thickness and the radius of each disk,
respectively, whereas $d$ is the inter-layer distance.}
\label{fig:ExpScheme}
\end{figure}

We test our model on a one-dimensional periodic stack of $N_d$ parallel disks randomly filled with $N_a$ atoms each, illuminated by a laser
beam incident under an angle $\theta_0$ with respect to the lattice axis (see Fig.~\ref{fig:ExpScheme}). We compare the predictions of the
microscopic model with those of the TM formalism noting that, while the TM approach assumes a radially infinite extension of the disks, our
model is able to account for any distribution, e.g., the Gaussian distribution common for thermal atomic clouds.

We emphasize that Eq.~(\ref{eq:stationary solution}) describes the scalar light scattering. We also present a full vectorial model in Section \ref{section:vectorial model}, giving the results in very good agreement with the scalar one derived for the one-dimensional lattice geometry.

\section{Bragg scattering.}
\begin{figure}[!ht]
\begin{tabular}{c}
\includegraphics[width=8cm]{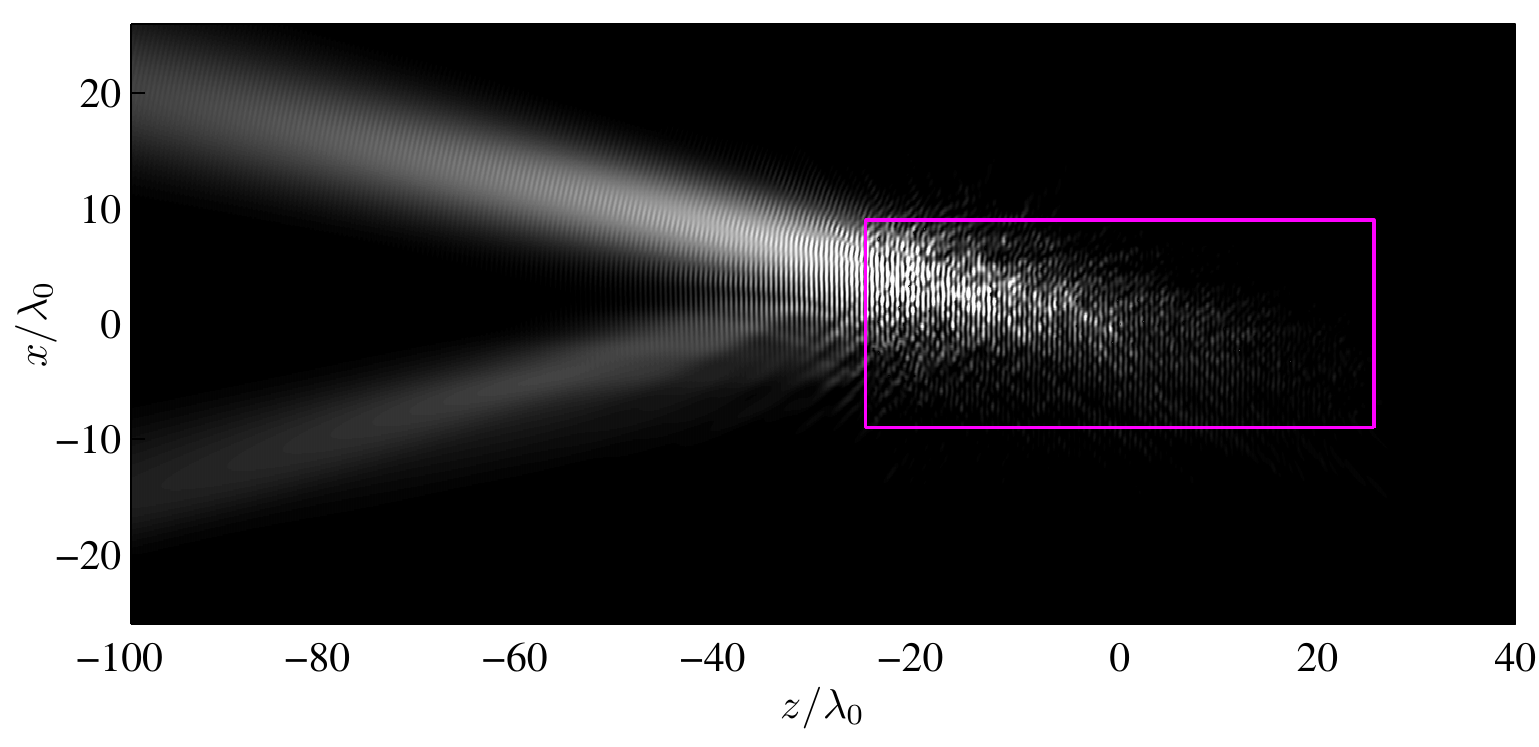}
\\ \includegraphics[width=8cm]{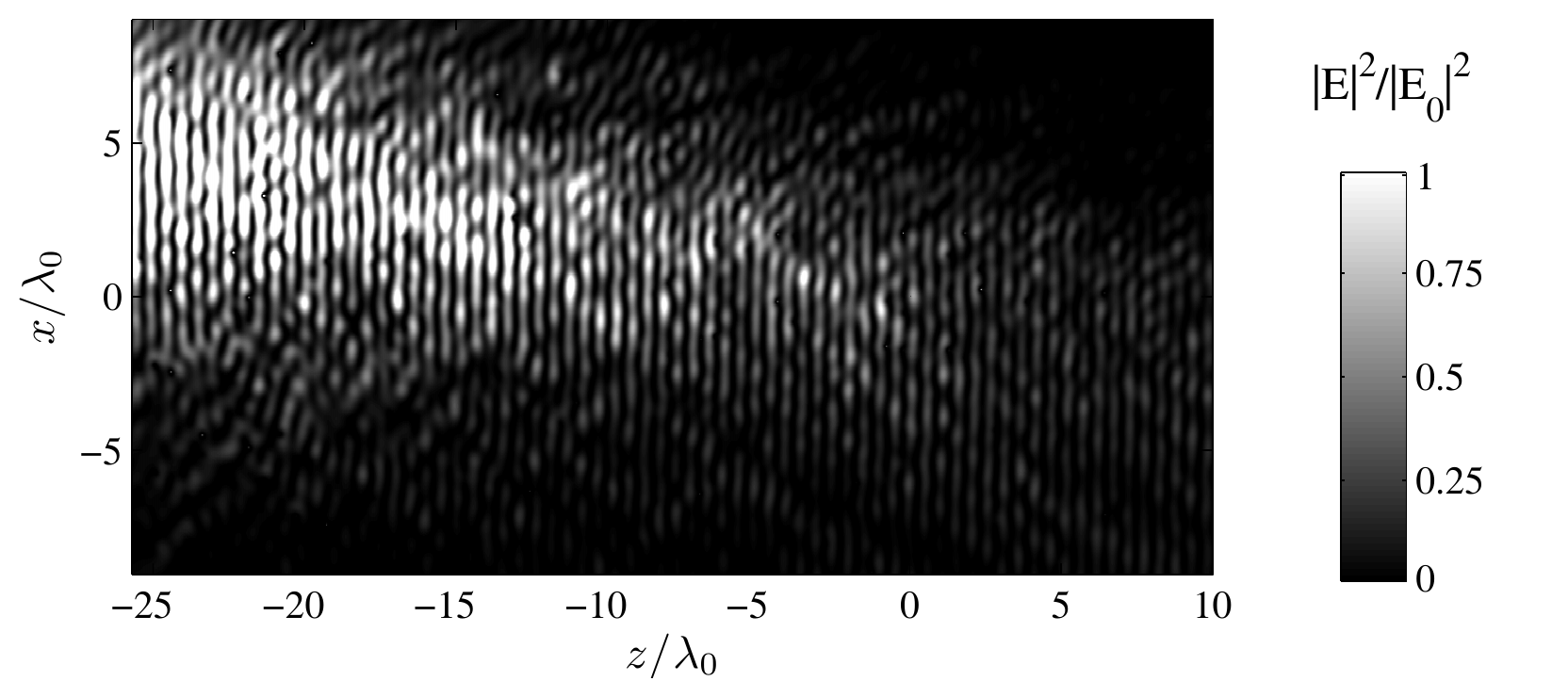}
\end{tabular}
\caption{(color online) The above picture: Intensity of the light in the $y=0$ plane as it enters a one-dimensional optical lattice and its reflection.
The rectangle marks the limit of the atomic structure. Below: Zoom of the left part of the atomic lattice. The luminous grains correspond to the strong
field radiated by the atoms close to the $y=0$ plane. The simulations are realized for $N=9000$ atoms randomly distributed over $N_d=100$ layers of
thickness $a=0.06\lambda_0$ and radius $R_d=9\lambda_0$, the distance between the atomic disks is $d=0.508\lambda_0$ with $\lambda_0$ being the
resonance wavelength. The Gaussian beam of waist $4.5\lambda_0$ and power $100~$mW is detuned by $\Delta_0=\Gamma$ from the atomic transition and
creates the angle $\theta_0=0.2~$rad with the lattice axis.} \label{fig:BraggScattering}
\end{figure}

We first investigate the scattering properties of our system under the Bragg condition, which means that the phase-shift of the incident wave
between two successive atomic disks is $\pi$. In this case, the interference of the waves reflected from each disk is constructive, and the
system is a Bragg reflector. This property is well reproduced by our model in which, despite the point-like nature of the scatterers, the
incident Gaussian beam is reflected by the atomic structure (see Fig.~\ref{fig:BraggScattering}), where we consider $^{85}$Rb atoms interacting
with the light fields via their $D2$ line. The total electric field $E$ is given by the sum of the incident field $E_0(\textbf{r})$ and the scattered
field
\begin{equation}    E_{scat}(\textbf{r})=-\frac{\hbar\Gamma}{\wp}\sum_j\beta_j\frac{\exp(ik_0|\textbf{r}-\textbf{r}_j|)}{k_0|\textbf{r}-\textbf{r}_j|},
\end{equation}
and according to the extinction theorem, the lattice produces in the forward direction a field opposed to the incident one. This demonstrates the
suitability of our model to study, e.g., the microscopic version of the Ewald–-Oseen theorem~\cite{Ballenegger1999}.

\begin{figure}[!ht]
\centerline{{\includegraphics[width=7cm]{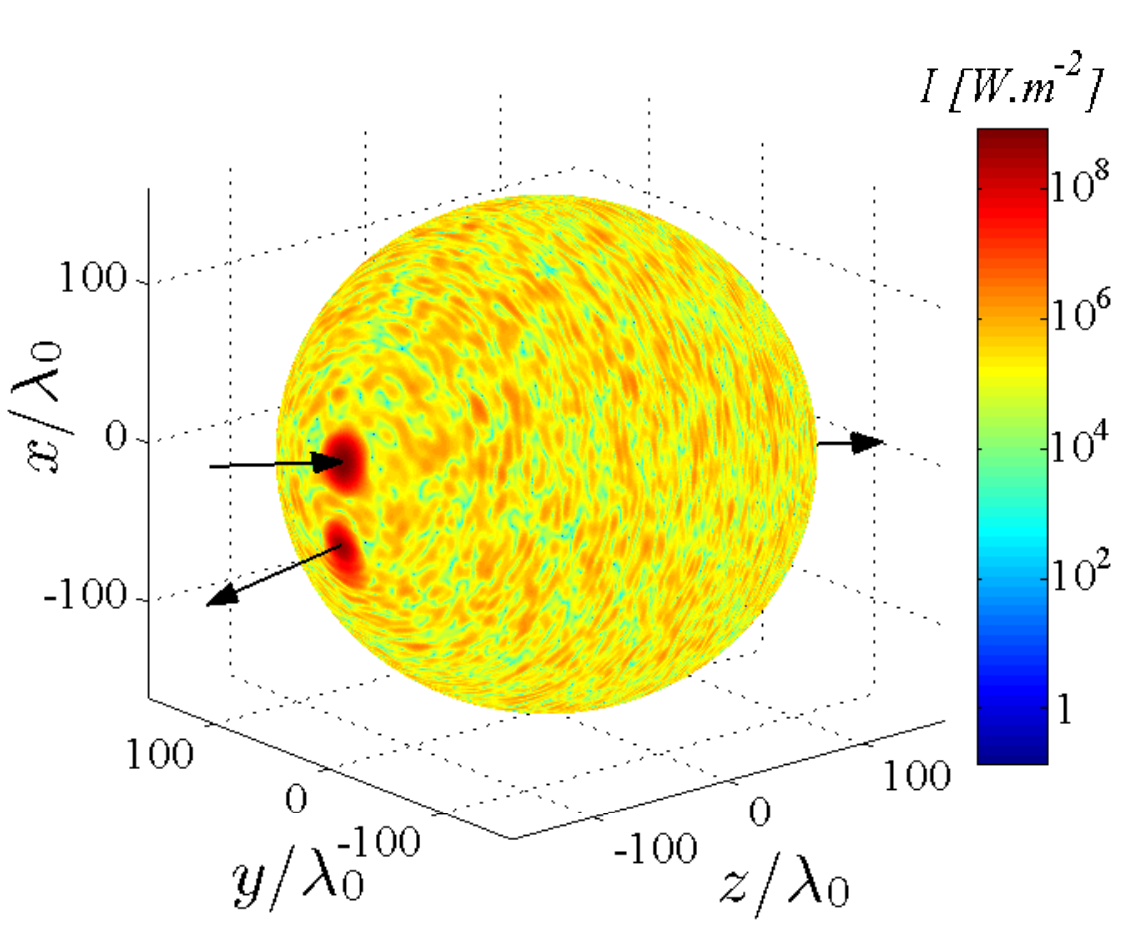}}}
\caption{(color online) Far-field intensity $I$ at the distance $150\lambda_0$ from the lattice. The light emitted into non-paraxial modes
exhibits a complex pattern because of the atomic disorder in the disks. The same parameters as for Fig.~\ref{fig:BraggScattering}
have been used.} \label{fig:SphereView}
\end{figure}

It can be observed that not all incident light is reflected by the
atomic structure. A significant part of it is re-emitted in the
form of the spontaneous emission. This phenomenon, which is
normally captured in the imaginary part of the refractive index,
is naturally present in the microscopic model (\ref{eq:stationary
solution}). The spontaneous emission appears in
Fig.~\ref{fig:SphereView} as the radiation into non-paraxial
modes. It should be noted that our microscopic model does not
contain light absorption, and we have verified that it conserves
energy, i.e., pursuant to the Maxwell's equations, the light which
is not reflected or transmitted is spontaneously scattered more or
less isotropically. The deviation from perfect isotropy, visible
in Fig.~\ref{fig:SphereView} as angular fluctuations, is a
signature of the disorder existing in each atomic disk.

\section{Photonic band gaps.}

\begin{figure}[!ht]
\centerline{{\includegraphics[width=8.5cm]{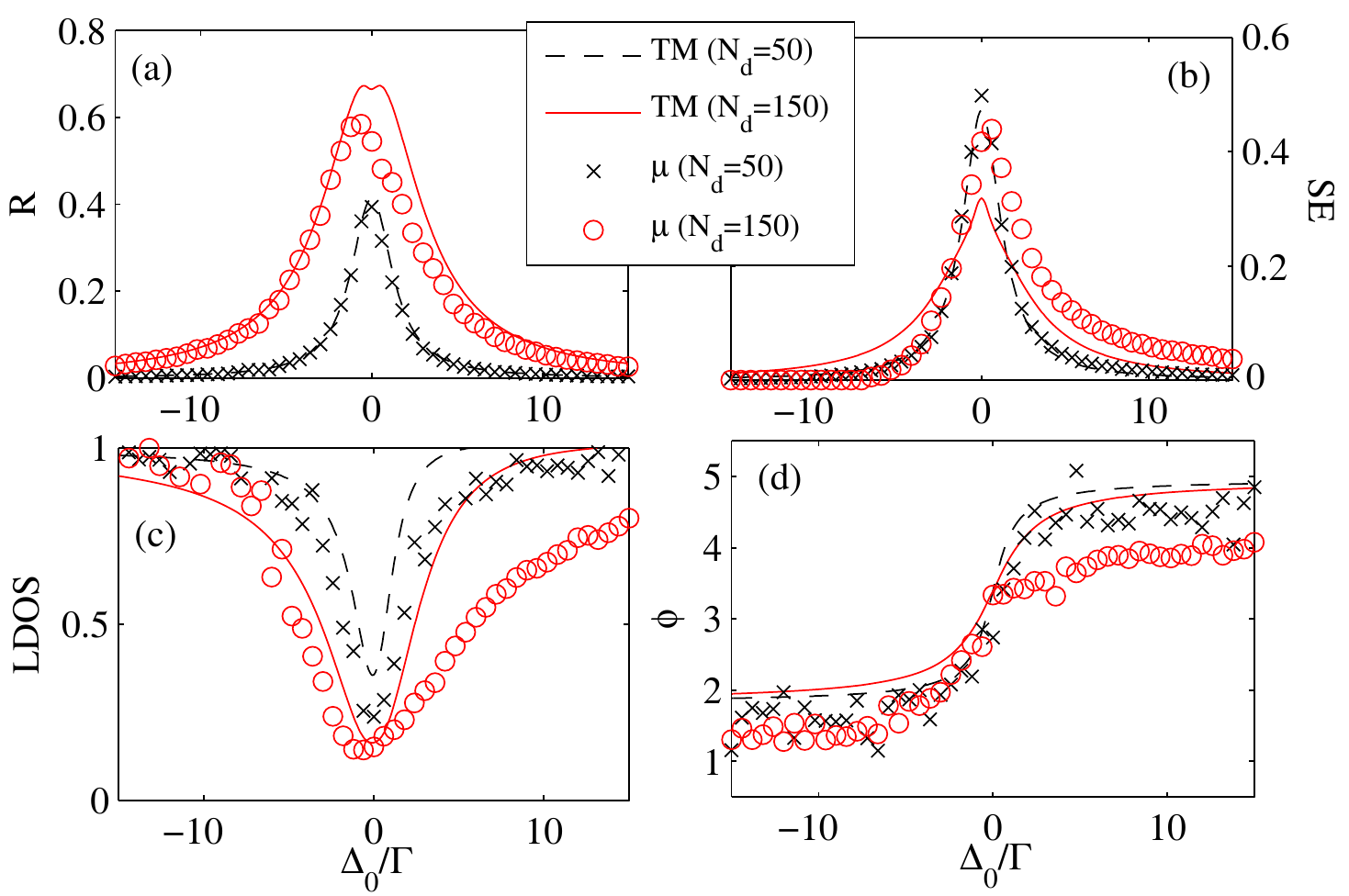}}}
\caption{(color online) Spectra of (a) the reflection coefficient $R$, (b) the spontaneous emission, (c) the LDOS and (d) the phase $\phi$ of the
complex reflection coefficient of the first half of the lattice. The simulations are realized for a semi-lattice with $N_d=50$ and $150$ layers of
thickness $a=0.04\lambda_0$ and the atomic density $\rho=5.95/\lambda_0^{3}$. The TM results correspond to the dashed black ($N_d=50$) and the
plain red ($N_d=150$) lines, whereas the symbols stand for the simulations of the microscopic model (black crosses and red circles, respectively).
The latter have been done for the atomic layers of radius $9\lambda_0$ filled with $N=3030$ and $9090$ atoms, respectively.}
\label{fig:PBG}
\end{figure}

Let us now turn to the study of band gaps. The
lattice reflectivity $R=|r_{N_{d}}|^2$ and the spontaneous
emission $SE=1-R-T$ (where $T=|t_{N_{d}}|^2$ is the lattice
transmissivity) in the microscopic model are in accord with the
predictions of the TM theory [see Figs.~\ref{fig:PBG}(a), 4(b)
and Eq.(\ref{rN})]. Here the reflection and transmission
coefficients are defined as $r_{N_{d}}=E_r/E_0$ and
$t_{N_{d}}=E_t/E_0$, where $E_{r,t}$ are the total electric fields
of the reflected and transmitted beam, respectively. Apart from a
high reflectivity, the presence of a PBG is characterized by the
local density of states (LDOS) vanishing. In the case of a
one-dimensional optical lattice, the LDOS at the center of the
lattice can be conveniently calculated using the complex
reflection coefficients $r_{-,+}$ of the two halves of the lattice
counting from the lattice center to its ends~\cite{Boedecker2003}:
\begin{equation}\label{eq:LDOS}
    D=\text{Re}\left(\frac{2+r_-+r_+}{1-r_-r_+}-1\right),
\end{equation}
taking into account that
$r_+=r_-e^{iN_d\cos\theta_0 k_0(a+d)}$. The
complex reflection coefficient $r_-=\sqrt{R_-}e^{i\phi}$ is
computed numerically using the reflectivity $R_-$ of the left
semi-lattice, given by the ratio of the reflected to the incident
power, and the phase $\phi$ of the wave reflected at the origin of
the lattice. On the one hand, $R_-$ has to be used to prevent
strong fluctuations in the local field $E_r(\mathbf{0})$ due to
the random distribution of the atoms arbitrarily close to the
origin; on the other hand, the phase $\phi$ needs to be taken at
the origin of the lattice to avoid extra phase-shifts that appear
at a distance in Gaussian beams (see the discussion below). As can
be seen in Fig.~\ref{fig:PBG}(c), a PBG is observed in our model,
which confirms the ability of this microscopic theory to capture
the photonic structure of the atomic cloud. However, the LDOS is
only in fair agreement with the results provided by the TM theory.
The discrepancy in the phase $\phi$ of the coefficient $r_-$ of up
to $\pi/4$ [see Fig.~\ref{fig:PBG}(d)] is explained by the fact
that our model treats the incident light as a Gaussian beam with a
finite waist and space-dependent phase shifts (e.g., the Gouy
phase), while the TM approach intrinsically assumes an incident
plane wave (see Section \ref{section:laser focalization}  for a discussion of the PBG properties on the laser focalization.). This affirms that Eq.~(\ref{eq:LDOS}) is reliable only
when the transverse finite-size effects are negligible.

Fig.~\ref{fig:PBG} also shows that the band gap appears as the number of disks is increased, and larger systems may exhibit deeper band gaps in their
spectrum. The simulation of large systems is actually the main limitation of our model since the complexity of the problem scales as $N^2$. However,
three-dimensional optical lattices typically contain several $10^4$ atoms \cite{Greiner2002}, rendering the computation feasible with standard computers.
This makes our microscopic model particularly promising in the quest of three-dimensional omnidirectional photonic band gaps in optical
lattices~\cite{Antezza2009,DeshuiYu2011}.

Finally, we have verified that under some idealizing hypotheses the microscopic model boils down to the TM formalism which is commonly used to
describe light propagation in one-dimensional atomic samples and characterize one-dimensional PBGs~\cite{Deutsch1995}.
The first step involves a coarse-graining of the atomic structure describing the atoms as a continuous density distribution $\rho(\textbf{r})$.
The atomic cloud is then characterized by a local refractive index $n(\textbf{r})=\sqrt{1-4\pi\rho/k_0^3(2\Delta_0/\Gamma+i)}$, and the wave
propagating in it can be shown to satisfy the Helmholtz equation~\cite{Bachelard2012}:
\begin{equation}
    \label{eq:Helmholtz}
    [\nabla^2+k_0^{2}n^2(\textbf{r})]E=0~.
\end{equation}
Furthermore, assuming the system to be one-dimensional, which in
practice means that its transverse size is much larger than the
lattice period and  the wavelength of the incident light, the
scattering problem can be reduced to a one-dimensional
wave-propagation problem and solved using classical techniques
such as the TM formalism~\cite{Born1999}. This explains the good
agreement of the latter approach with our microscopic theory up to
the point, where finite-size effects start playing a significant
role [see Fig.~\ref{fig:PBG}]. According to our derivation, the
reflection and transmission coefficients $r_{N_d}$ and $t_{N_d}$
for an atomic structure consisting of $N_d$ parallel disks
of uniform density can be written in terms of the
reflection and transmission coefficient amplitudes $r$ and $t$ for a single layer and the Bloch phase
$\phi_B$:
\begin{eqnarray}
    r_{N_d}&=&\frac{r\sin N_d\phi_B}{\sin
    N_d\phi_B-t\sin(N_d-1)\phi_B}\label{rN}\\
    t_{N_d}&=&\frac{t\sin\phi_B}{\sin
    N_d\phi_B-t\sin(N_d-1)\phi_B}~,\label{tN}
\end{eqnarray}
 where $\phi_B$ is determined by
 \begin{equation}\label{phiB}
    \cos\phi_B=\cos(k_{0z}d)\cos(k_{z}a)-\frac{k_{0z}^2+k_{z}^2}{2k_{0z}k_{z}}\sin(k_{0z}d)\sin(k_{z}a)\nonumber
\end{equation}
with $k_{0z}=k_0\cos\theta_0$ and
$k_z=k_0\sqrt{n^2-\sin^2\theta_0}$. The detailed derivation of
these results which are consistent with other models~\cite{Deutsch1995,Bendickson1996} will be reported elsewhere.

\section{Vectorial model.} \label{section:vectorial model}
In this section  we provide the full vectorial model which is necessary to describe higher dimensions, and its comparison with the scalar one in one-dimensional geometry.

In the absence of degeneracy of the excited state, the interaction of an ensemble of two-level atoms with vectorial light is described by the following equation~\cite{Manassah2012}:
\begin{eqnarray}\label{betajk}
    \left(i\Delta_0-\frac{\Gamma}{2}\right)\beta_j^{(\alpha)}&=&
    \frac{i\wp}{2\hbar}E_0^{(\alpha)}(\textbf{r}_j)
    \\ &&+\frac{\Gamma}{2}\sum_{\alpha'}\sum_{m\neq j}G^{(\alpha,\alpha')}(\textbf{r}_j-\textbf{r}_m)\beta_m^{(\alpha')}\nonumber
\end{eqnarray}
with $\alpha=(x,y,z)$, and the vectorial kernel reads
\begin{eqnarray}\label{Gjm}
G^{(\alpha,\alpha')}(\mathbf{R})&=&\frac{e^{ik_0R}}{ik_0R}
\Bigg[\delta_{\alpha,\alpha'}-\hat R^{(\alpha)}\hat R^{(\alpha')}
\\ &&+ \left(\frac{i}{k_0 R}-\frac{1}{k_0^2 R^2} \right)
(\delta_{\alpha,\alpha'}-3\hat R^{(\alpha)}\hat R^{(\alpha')}) \Bigg],\nonumber
\end{eqnarray}
where $\mathbf{R}=R\hat
{\mathbf{R}}$. The vectorial electric field components of the scattered wave at point $\mathbf{r}$ can be written as a function of $\beta_j^{(\alpha)}$ as
\begin{equation}\label{Ebeta}
E_{scat}^{(\alpha)}(\mathbf{r})=-\frac{i\hbar\Gamma}{\wp}\sum_{\alpha'}\sum_j G^{(\alpha,\alpha')}(\textbf{r}-\textbf{r}_j)
\beta_j^{(\alpha')}.
\end{equation}

\begin{figure}[!ht]
\begin{tabular}{cc}
\includegraphics[width=4cm]{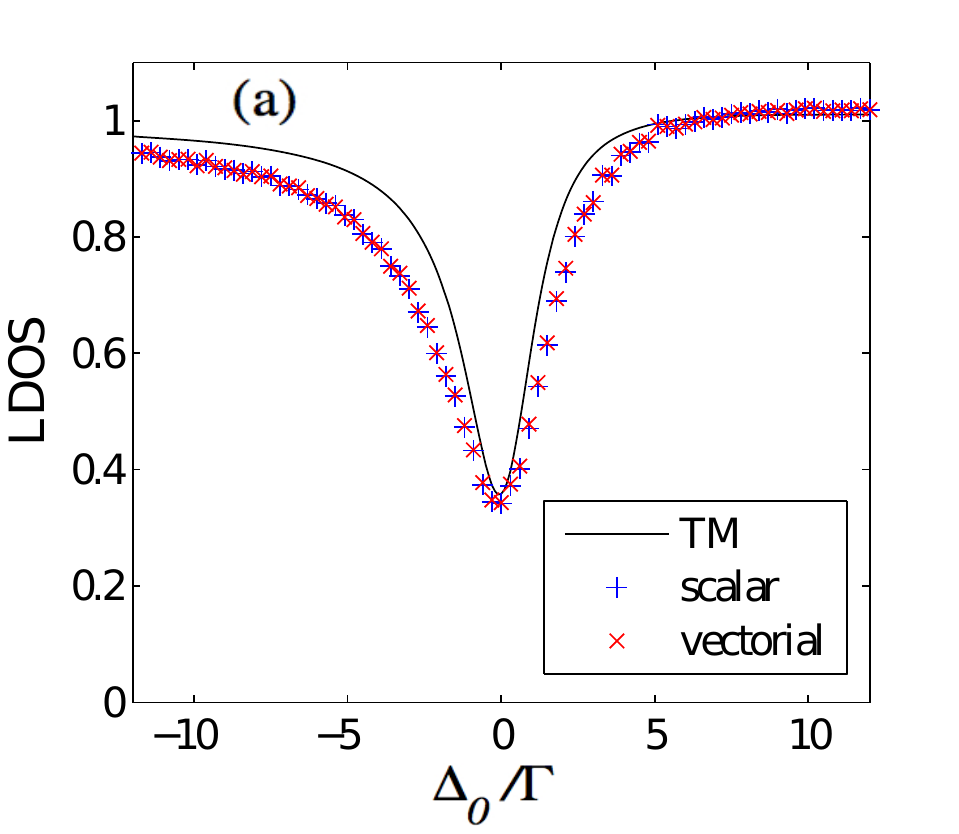}& \includegraphics[width=5cm]{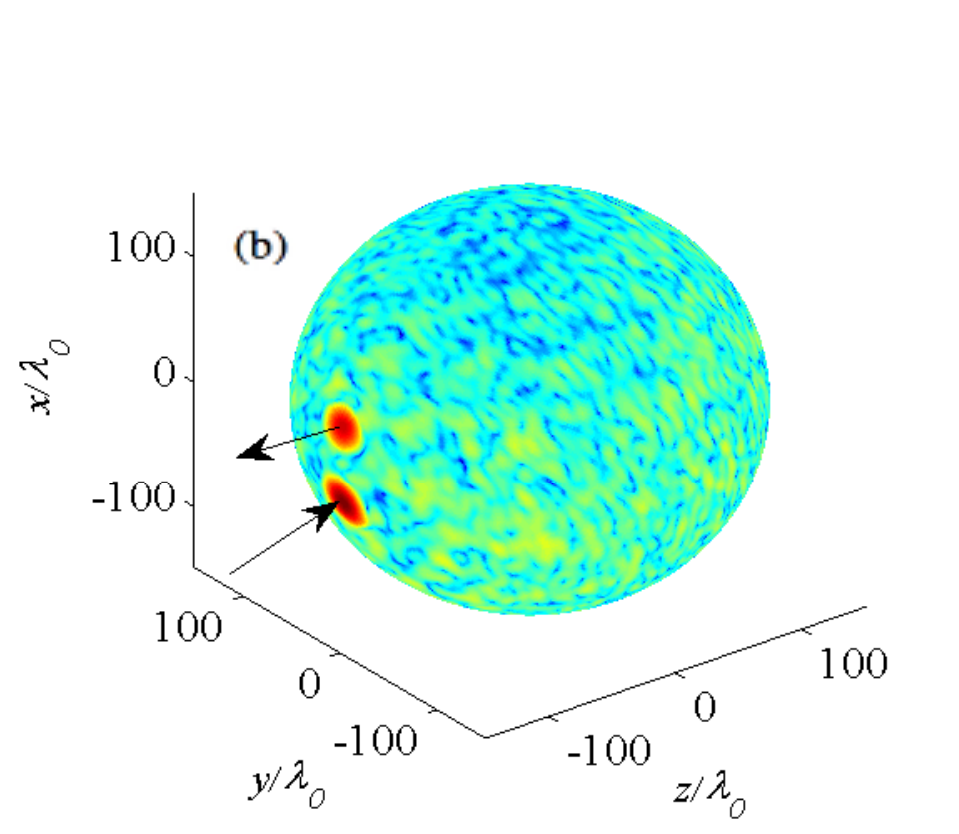}
\end{tabular}
\caption{(color online) Left: LDOS derived from the TM formalism (black plain curve), the scalar model (blue pluses) and the vectorial model (red crosses). The agreement between the results provided by the two microscopic models is excellent. The simulations are realized for $N=3000$ atoms spread over $N_d=40$ disks of radius $R_d=9\lambda_0$ and thickness $a=0.05\lambda_0$, the  angle of incidence is $\theta_0=0.035$ rad. Right: Far-field intensity $I$ at the distance $150\lambda_0$ from the lattice. The same parameters as for the left picture have been used; the detuning is $\Delta_0=0$ and the incident light is linearly polarised along the $x$ axis. The vectorial model reveals the anisotropy of the spontaneous emission pattern, depending on the polarization of the incoming light.\label{fig:vectorial}}
\end{figure}

In the one-dimensional geometry discussed in this letter, the vectorial model predicts the opening of a PBG for the same set of parameters as for the scalar model. Indeed, as can be observed in Fig.~\ref{fig:vectorial}(a), the LDOS derived from both models are in perfect agreement. The vectorial nature of the incident light really comes into play if one focuses on the structure of the spontaneous emission (see Fig.~\ref{fig:vectorial}(b)) or in the case of two- or three-dimensional optical lattices.

\section{Laser focalization.}\label{section:laser focalization}

\begin{figure}[!ht]
\begin{tabular}{c}
\includegraphics[width=8cm]{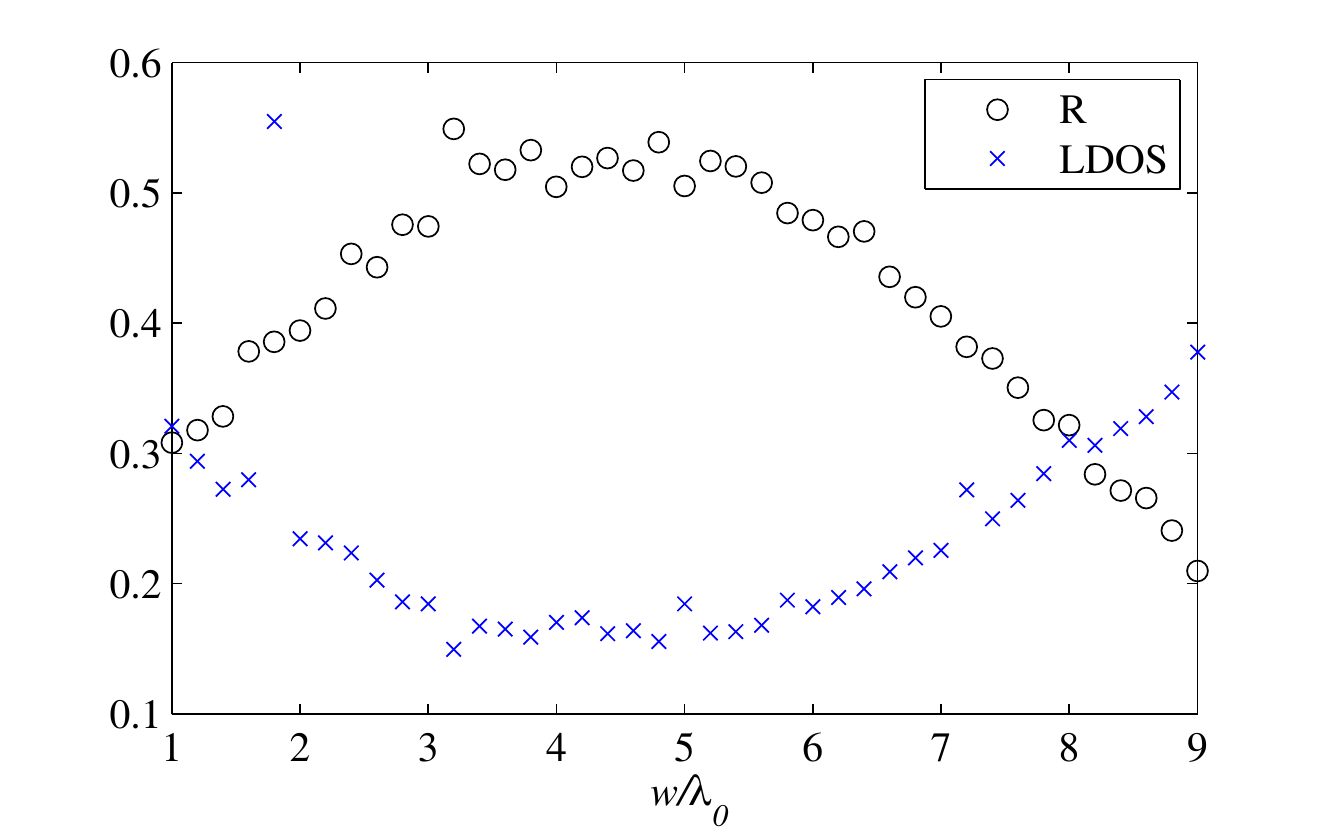}
\end{tabular}
\caption{(color online) Reflection coefficient (dark circles) and LDOS (blue crosses) derived from the scalar model as the laser waist is tuned. The simulations are realized for $N=6000$ atoms distributed over $N_d=60$ disks of radius $R_0=9\lambda_0$ and thickness $a=0.05\lambda_0$; the laser is incident under the angle $\theta_0=0.035$ rad and is at resonance, i.e., $\Delta_0=0$.\label{fig:waist}}
\end{figure}
The TM formalism describes the scattering of a plane-wave from a lattice of infinite transverse size. However, finite-size effects such as the laser divergence may alter the results of the TM theory. Our numerical simulations realized for the disks of radius $R_d=9\lambda_0$ have revealed that the PBG is optimized when the waist of the incoming laser is approximately half of that radius. Indeed, Fig.~\ref{fig:waist} shows that the reflectivity reaches its maximum value and the LDOS is minimal for a laser waist in the range of $3-6\lambda_0$. Due to the finite size of the lattice, large waist lasers will not be fully intercepted by the lattice, while for small waists the divergence of the laser makes the photons meet a lower number of disks, thus reducing the PBG.

\section{Conclusion.}

In conclusion, we have proposed a microscopic description of the scattering of light from optical lattices and demonstrated how multiple
reflections from adjacent lattice sites can open a photonic band gap. The reconsideration of photonic bands under the microscopic scattering
perspective leads not only to a deeper understanding of the phenomenon, but also offers a practical advantage of a larger range of applications
than other models. For instance, our model naturally includes the cloud's granularity. Defects, such as site vacancies or finite-size effects,
can be easily taken into account, making the microscopic model particularly promising to study the transition from ordered to disordered clouds.
This feature distinguishes our model from approaches based on the expansion of the electric field in terms of Bloch wave vectors and the solution
of the Helmholtz equation in reciprocal space~\cite{Antezza2009,DeshuiYu2011}. Those models require optical lattices to be infinite and quasi-periodic.

Moreover, in comparison to the transfer matrix approaches \cite{Deutsch1995}, the microscopic model can be readily extended to two- or
three-dimensional lattices of arbitrary geometry. It also allows for the description of experimental side effects in one-dimensional lattices,
such as those considered in this paper, e.g., walk-off losses~\cite{Slama2005,Schilke2012} as the result of the finite radial extension of the
atomic disks, the deviation of the probe laser beam entering the atomic cloud caused by the refraction and the inhomogeneous Stark-shift due to
the intensity distribution of the trapping light \cite{Slama2006}.

The price to pay is that numerical simulations get quite heavy beyond a few $10^4$ atoms, which however, is not too far from experimentally
relevant systems. These atomic numbers are proven to be sufficient enough to reach Bragg scattering and important reflection coefficients
comparable to those obtained experimentally \cite{Schilke2011}. Hence, we believe the microscopic model to be appropriate to characterize
other collective phenomena usually approached using coarse-grained theories, where the medium is described by a refractive index.
In the case of disordered systems, the proposed model paves the way for a microscopic discussion of the extinction theorem~\cite{Ballenegger1999}
and the Abraham-Minkowski controversy~\cite{Barnett2010}.

Finally, it is worth noticing that, despite the fact that multiple scattering is naturally included into microscopic collective scattering models,
the major part of recent studies avoid this regime, where the physical interpretation of the observed effects can be ambiguous. In contrast, our
work points out that ordered lattices represent systems, where multiple scattering leads to the relatively simple and well-known phenomenon of
photonic band gaps.

During the preparation of our paper we became aware of
another work on photonic band gaps by M. Antezza and Y. Castin, arXiv:1304.7188, where
the authors investigate finite-size effects in diamond lattices.

\section{Acknowledgments.}

This work has been supported by the Funda\c{c}\~{a}o de Amparo \`{a}
Pesquisa do Estado de S\~{a}o Paulo (FAPESP) and the Research
Executive Agency (Program COSCALI, Grant No. PIRSES-GA-
2010-268717).

\end{document}